\documentclass[preprint,twocolumn]{aastex7}

\newcommand{\swiftnosp}{\mbox{\textit{Swift}}}
\newcommand{\BAT}{\mbox{\textit{Swift}-BAT~}}

\newcommand{\cgro}{\mbox{\textit{CGRO}-BATSE~}}

\newcommand{\rxte}{\mbox{\textit{RXTE}-PCA~}}
\newcommand{\rxtenosp}{\mbox{\textit{RXTE}-PCA}}

\newcommand{\fermi}{\mbox{\textit{Fermi}-GBM~}}

\newcommand{\integral}{\mbox{\textit{INTEGRAL~}}}
\newcommand{\integralnosp}{\mbox{\textit{INTEGRAL}}}

\usepackage{comment}
\usepackage{amsmath, soul}
\usepackage{multirow}

\shorttitle{Magnetar Age and Bursts}
\shortauthors{Keskin et al.}

\begin{document}

\title{Correlation of Burst Behaviour with Magnetar Age}

\correspondingauthor{\"Ozge Keskin}
\email{ozgekeskin@sabanciuniv.edu}

\author[0000-0001-9711-4343]{\"Ozge Keskin}
\affiliation{Sabanc\i~University, Faculty of Engineering and Natural Sciences, \.Istanbul 34956 T\"urkiye}
\email{ozgekeskin@sabanciuniv.edu}

\author{Samuel K. Lander}
\affiliation{School of Engineering, Mathematics and Physics, University of East Anglia, Norwich, NR4 7TJ, U.K. }
\email{Samuel.Lander@uea.ac.uk}

\author[0000-0002-5274-6790]{Ersin G\"o\u{g}\"u\c{s}}
\affiliation{Sabanc\i~University, Faculty of Engineering and Natural Sciences, \.Istanbul 34956 T\"urkiye}
\email{ersing@sabanciuniv.edu}


\begin{abstract}

We analyze a wide set of historical magnetar burst observations detected with five different instruments, calibrating these to the energy range of \fermi observations for consistency. We find a striking correlation between a magnetar's characteristic age and both its typical burst energy and its burst activity level. Arguing that this bursting behaviour also correlates with true age, we interpret it as the result of a reducing high-stress volume of the crust in an aging magnetar: previous giant flares cause relaxation of large regions of its crust and inhibit burst clustering, whilst the reducing burst energy reflects the progressively shallower region of the crust where Hall drift can build stresses effectively, as the field decays through the range $\sim 10^{12}-10^{13}\,\mathrm{G}$. Low-energy bursts from very young magnetars may represent failures of weak regions of the crust that have only recently solidified.
\end{abstract}

\keywords{\uat{High Energy astrophysics}{739} --- \uat{Neutron Stars}{1108} --- \uat{Magnetars}{992} --- \uat{X-ray bursts}{1814}}


\section{Introduction} \label{sec:introduction}

One of the most notable and distinctive features of the highly magnetized isolated neutron stars known as magnetars \citep{DT92} is their recurring emission of brief but highly luminous bursts in hard X-rays or soft $\gamma-$rays. These bursts, typically lasting $\sim$\,0.1 seconds, release immense amounts of energy, reaching up to 10$^{42}$ erg \citep{EG01}. Around two-thirds of known magnetars have exhibited such bursts at various occurrence rates \citep{Olausen2014}. During a burst active phase of a magnetar, between a single to thousands of bursts may occur, with activity episodes lasting as long as several months. 

Short magnetar bursts are generally attributed to the local yielding of the solid neutron star crust resulting from the release of accumulated elastic stress due to the evolution of the star's internal magnetic field $B$ \citep{TD95, TD01}. This energy is transferred to the magnetosphere, where a burst may be produced as a result of a magnetic reconnection event, i.e., a rapid untwisting of highly twisted external magnetic field lines \citep{TD95, lyu03}.

In some magnetars, the emission of energetic burst(s) also marks the onset of `magnetar outbursts': enhanced persistent X-ray emission episodes lasting from weeks to years. In their extensive study of X-ray observations of outbursts from 17 magnetars, \cite{CotiZelati} showed that the total energy released in each outburst event is inversely correlated with the characteristic age of the star, as determined from its spin period and spindown rate. In other words, they reported that young magnetars are more likely to exhibit energetic outbursts than older ones. They attributed this behavior to magnetic field decay, which limits the available energy budget as a magnetar ages.

Similarly, when looking at typical short-burst emission, magnetars with low characteristic ages are burst prolific \citep[see e.g.,][]{van12, Lin2020} and often exhibit clusters of bursts \citep{Kaneko10, Kaneko21}. In this paper, we perform a systematic investigation of short burst activity behavior in magnetars by utilizing numerous observational studies of magnetar bursts and long-term monitoring of their spin (Section \ref{sec:observation}), finding an important correlation between burst behaviour and age (Section \ref{sec:result}), which we show can be used to help understand the evolving stress pattern in a magnetar's crust (Section \ref{sec:theory}). We then discuss the implications of our study in Section \ref{sec:discussion}.


\section{Sample} \label{sec:observation}

We aim to characterize typical short bursts emitted by numerous Galactic magnetars over the past three decades. Specifically, we compiled burst fluences from time-integrated spectral studies in the literature. Using distance estimates \(D\) for each burst-emitting source, we calculated burst energetics under the assumption of isotropic energy release. Additionally, we gathered the spin period \(P\) and its rate of change \(\dot{P}\) for each source, allowing us to determine the characteristic ages \(\tau_c\) of the magnetars in our sample and their inferred dipolar magnetic field strengths ($B$). We listed the compiled database in a table in Appendix \ref{appA}. We also extended a discussion about the distances to sources utilized in this paper in Appendix \ref{appB}.

The compiled burst characteristics are based on analyses of data from five different hard X-ray telescopes. To ensure consistency across all burst properties, we calibrated the fluences by accounting for the sensitivity (detector response) of each telescope. We assume the most characteristic spectral shape of short magnetar bursts, which is a power law with exponential cutoff whose $\nu$F$_\nu$ spectrum peaks at an energy $E_{\rm peak}$ \citep{Lin2011, van12}. We take model parameters of $E_{\rm peak} = 35$ keV and photon index $\Gamma=0$ \citep{Collazzi2015,Lin2016,Lin2020}. We then obtained conversion factors using Xspec (version 12.14.1) to calibrate fluence values calculated in different energy bands to the energy range of \fermi observations, since most of the bursts observed in the last two decades were detected with it. Below, we detail the specific considerations for bursts detected with each of these five instruments.


\begin{figure*}[htbp!]
    \centering
    \includegraphics[width=0.775 \textwidth, trim=45 260 60 235, clip]{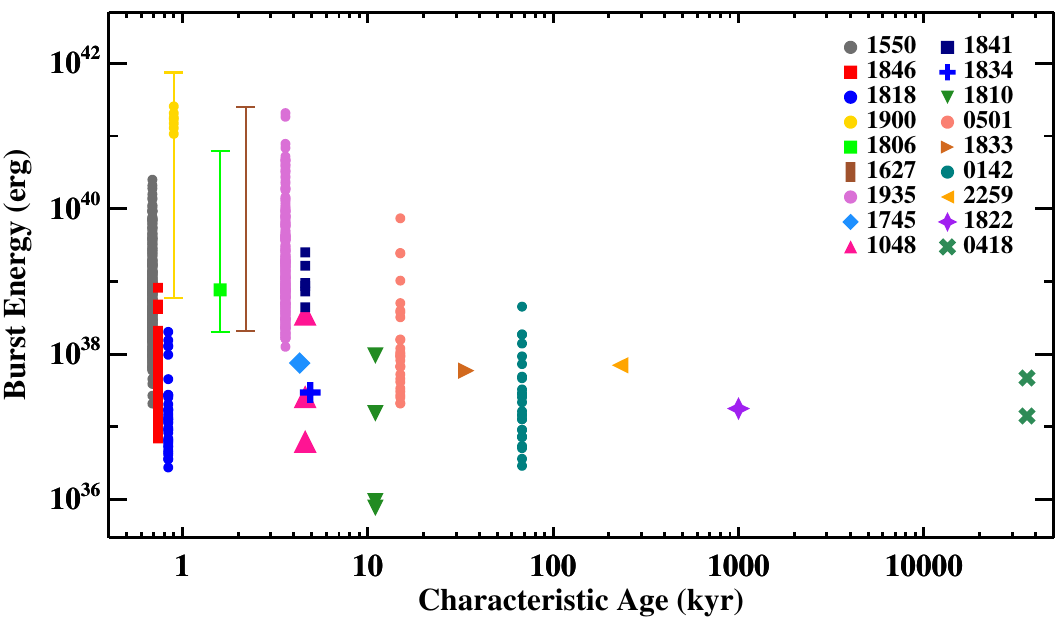}
    \caption{Plot of burst energies vs. characteristic ages of burst active magnetars. Filled symbols represent the energetics of individual bursts from the color-coded magnetars. For SGR 1900+14 (gold), SGR 1806$-$20 (green), and SGR 1627$-$41 (brown), we present burst energy ranges rather than individual burst energies (see text for details). Symbols on the top right are ordered in increasing characteristic age.}
    \label{fig:energy_vs_age}
\end{figure*}


\textbf{\fermi Observations:} The sample from this telescope includes the entire burst observations of SGR 0418+5729, SGR 0501+4516, SGR J1550$-$5418, Swift J1818.0$-$1607, Swift J1822.3$-$1606, 1E 1841$-$045, PSR J1846.4$-$0258, SGR J1935+2154, and 1E 2259+586. Moreover, it also includes short bursts of AXP 4U 0142+61 in February 2015 and a single SGR 1806$-$20 burst in March 2010. Note that some bright bursts from SGR J1550$-$5418 \citep{van12} and SGR 0501+4516 \citep{Lin2011} exceeded the instrumental readout capabilities and caused data saturation. For these cases, we obtained the isotropic burst energies from the non-saturated time segments of those bright bursts. The fluences (therefore, energies) of \fermi detected bursts were reported in the energy range of 8$-$200\,keV.

\textbf{\BAT Observations:} Short bursts from AXP 4U 0142+61 were observed with \BAT in July 2011 and a single burst in June 2012. Additionally, single bursts from three other sources, SGR J1745$-$2900, SGR J1833$-$0832, and Swift J1834.9$-$0846, were detected with \swiftnosp. Their reported fluences were in the energy range of 15$-$150 keV. To calibrate these values to 8$-$200 keV, we obtained a conversion factor of 1.17 by following the approach described earlier.
Moreover, the SGR 1900+14 burst forest in March 2006 was detected with \swiftnosp. Spectral analysis of the seven slightly longer duration events during the forest was performed in the energy range of 10$-$100 keV \citep{Israel08}; for these, we assumed an average burst duration of 1 s, and determined a conversion factor of 1.07 to obtain fluence values in 8$-$200 keV.

\textbf{\cgro Observations:} For the bursts of SGR 1900+14 in 1998-1999 and SGR 1806$-$20 in 1993, 1996, and 1998 observed with \cgro, the energy fluences were reported in the energy range of $>25$\,keV, hence, only the minimum energy bound of the integration energy interval is known. Similarly, for the bursts from SGR 1627$-$41 in 1998, the peak energy flux values were in $>25$\,keV \citep{Wood1999}. The upper energy bound for fluences (and fluxes) is effectively 200\,keV as the spectrum drops exponentially above $E_{\rm peak}$. We assumed the durations of bursts as 0.1 s, which is the typical burst duration of their sample. Therefore, for the fluences of bursts observed with \emph{CGRO} ($>25$\,keV), we obtained a conversion factor of 1.59 to calibrate them to 8$-$200 keV. Finally, we note that, for these observations, the energy fluence/flux ranges of the burst active episodes (i.e., the minimum and maximum burst fluence/flux values) rather than the individual ones were reported.

\textbf{\integral Observations:} The intense burst active episode of SGR 1806$-$20 in 2003-2004 was observed with INTEGRAL. \cite{Gotz2006} analyzed the data of 224 bursts during this period in the 15-100 keV range. We estimated the energy fluence range of these bursts by utilizing their fluence histogram; hence, similar to \cgro observations, we have only the burst fluence range instead of having individual fluences. We calculated a conversion factor of 1.2 for the calibration of fluences in the 15$-$100 keV band to 8$-$200 keV.

\textbf{\rxte Observations:} The burst activities from two sources, 1E 1048.1$-$5937 and XTE J1810$-$197, presented in this study were observed with \rxtenosp. The temporal and spectral analyses of these events were performed in 2$-$20/30\,keV \citep{Gavriil2002, Gavriil2006, Wood2005}. The burst profiles of these sources are quite similar to one another, yet different from those of other sources. As for temporal characteristics, they exhibit longer durations, characterized by a short spike followed by a long tail. Due to its proximity to SGR 1806$-$20, the identification of XTE J1810$-$197 as the source of the four bursts observed in 2003-2004 was only possible through the detection of 5.54-second pulsations in the several-hundred-second-long tails of the events \citep{Wood2005}. As for spectral characteristics, they have softer spectra that can be modeled with a single blackbody with temperatures around a few keV. That is why the energetics of these events are lower even after applying the integration energy interval calibration\footnote{We employed a single blackbody model with the reported blackbody temperatures of a few keV, which results in significant emission below 8 keV. Therefore, we calibrated the energies within 2$-$200 instead of 8$-$200\,keV.}. We, therefore, keep these bursts out of our sample for further analysis due to their significant spectral and temporal differences. However, including or excluding them does not significantly affect our findings. Nevertheless, we still plot them in our figures and note that the reason they have lower yet comparable energetics to regular short and bright magnetar bursts is their long durations due to X-ray bright tails.

We present magnetar-based burst energies, calibrated as discussed above, vs. their calculated characteristic ages in \autoref{fig:energy_vs_age}. Note again that for the activities observed with \cgro and \integral from the sources, SGR 1900+14 (gold), SGR 1806$-$20 (green), and SGR 1627$-$41 (brown), we present burst energy ranges.


\section{Results} \label{sec:result}

\begin{figure*}[htbp!]
    \centering    \includegraphics[width=0.775 \textwidth, trim=45 260 60 235, clip]{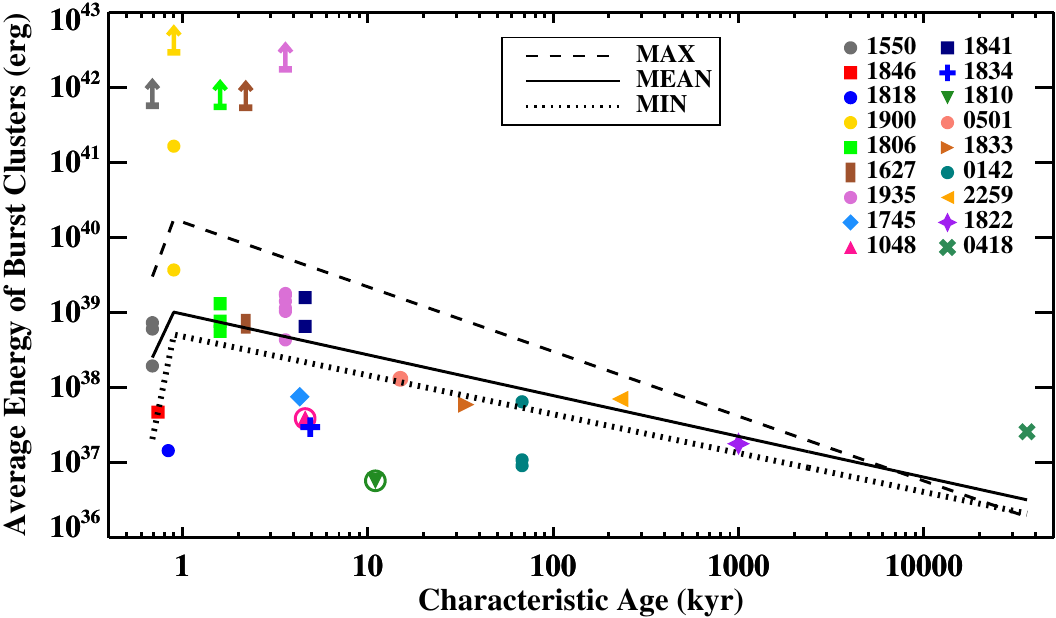}
    \caption{Plot of mean energies of burst clusters vs. characteristic ages of burst active magnetars. Filled symbols represent the mean energetics of burst active episodes of the color-coded magnetars. The solid line shown represents the best BPL fit to the data, with dashed and dotted lines showing BPL fits to the maximum and minimum burst energies of clusters, respectively. The activities of 1E 1048.1$-$5937 and XTE J1810$-$197, represented by upward and downward triangles inside circles, were not included in the fits (see the text for details). Upward arrows display the total burst energies considering all activities of the sources in our sample. Symbols on the top right are ordered in increasing characteristic age.}
    \label{fig:burst_clus_vs_age}
\end{figure*}

We observe a relation between the magnetar's burst energy and characteristic age, as seen in \autoref{fig:energy_vs_age}: Older magnetars emit few bursts with low energies, while younger magnetars exhibit many bursts (often in clusters) with higher energies. To quantify this relation, we fit the trend as follows: we first computed the mean burst energy for each burst active episode of each source, considering the burst active episodes/burst clusters separately (see \autoref{fig:burst_clus_vs_age}). Any cluster size may range from a single burst activity to hundreds of bursts. Our motivation for using burst clusters rather than individual bursts in this analysis was to allow a fairer comparison between sources, rather than deriving a trend heavily skewed by prolific young bursting sources like SGR 1935+2154.


For the burst active episodes of three sources, observed with \emph{CGRO} and \integralnosp, we estimated mean energies by utilizing the reported energy fluence/flux distributions. 
Guided by the broad trend of the data, as well as theoretical expectations (see the following section), we then fit the mean energy values with a Broken Power Law (BPL) model (see the solid line in \autoref{fig:burst_clus_vs_age}). The best fit yields a break $\tau_{\rm break}$ at a characteristic age of 0.9 kyr, with the power-law indices before and after the break being 5.40 and $-$0.54, respectively.


\begin{deluxetable*}{lcccccr}[htbp!]
\tablecaption{Power-law (PL) fits to our data for burst energy versus $\tau_c$. We show the pre- and post-break PL indices $\alpha,\beta$ assuming a broken PL model, and the index $\gamma$ if the data are fitted to a single PL.}
\tablewidth{0pt}
\tablehead{
& & & \colhead{\bf BPL}  & & \colhead{\bf PL} \\ 
\cline{3-5}
& & \colhead{$\tau_{\rm break}$ (kyr)} &  \colhead{$\alpha$} & \colhead{$\beta$} & \colhead{$\gamma$}}
\startdata
\hline
\multirow{3}*{\parbox[c]{12mm}{\centering
    Per Burst Cluster}} & Min &  0.90 $\pm$ 0.11 & 12.33 $\pm$ 7.39 & $-$0.52 $\pm$ 0.20 & $-$0.35 $\pm$ 0.19 \\
    \cline{2-6}
    & Mean &  0.90 $\pm$ 0.19 & 5.40 $\pm$ 5.95 & $-$0.54 $\pm$ 0.16 & $-$0.48 $\pm$ 0.14 \\
    \cline{2-6}
    & Max & 0.90 $\pm$ 0.22 & 6.62 $\pm$ 9.13 & $-$0.88 $\pm$ 0.25 & $-$0.78 $\pm$ 0.21\\
    \hline
    \multirow{3}*{{\parbox[c]{14mm}{\centering
    Per Source}}} & Min & 0.96 $\pm$ 0.19 & 8.90 $\pm$ 7.23 & $-$0.28 $\pm$ 0.15 & $-$0.13 $\pm$ 0.14 \\
    \cline{2-6}
    & Mean &  0.94 $\pm$ 0.38 & 4.34 $\pm$ 8.18 & $-$0.37$\pm$ 0.17 & $-$0.30 $\pm$ 0.14 \\
    \cline{2-6}
    & Max & 0.90 $\pm$ 0.41 & 6.06 $\pm$ 15.61 & $-$0.77 $\pm$ 0.33 & $-$0.69 $\pm$ 0.26\\
    \hline
\enddata
\end{deluxetable*}\label{tab:fit_results}


In case of a BPL fit to the minimum energies of each burst cluster (see dotted line in \autoref{fig:burst_clus_vs_age}), the slopes change to 12.33 and $-$0.52, respectively, with a break at the same characteristic age. When we fit the maximum burst cluster energies with a BPL (dashed line in \autoref{fig:burst_clus_vs_age}), we obtain the same break value with indices of 6.62 and $-$0.88, respectively.

To check possible alternative correlations, we also fitted the minimum, mean, and maximum energies of each burst cluster with a Power Law (PL) model. Finally, we obtained an average burst energy for each source, instead of averaging them per burst cluster, and fitted this data (i.e., minimum, mean, and maximum per-source burst energies) to both PL and BPL models.

We list the results of all these fits in \autoref{tab:fit_results}. Fit results are consistent with one another within their 3$\sigma$ errors. Also, the errors of BPL indices $\alpha$ before $\tau_{\rm break}$ are large due to the deficiency of data points before the break (i.e., the few magnetars younger than $\tau_{\rm break}$). By contrast, since many data points lie after the breaks of BPLs, the PL indices $\beta$ for these are better constrained. For the same reason, their values are also compatible with the indices $\gamma$ of simple PL fits.


\section{Theoretical interpretation}\label{sec:theory}

\subsection{Physical differences between magnetars}

Although a number of attributes of a magnetar can be inferred, very few are genuine observables that are common to a number of sources. Our results show a correlation involving three directly observed parameters of burst-active magnetars: their spin period and its time derivative (these two combined into the characteristic age), and their burst activity. Uncertainties in the physics of magnetars allow for various different scenarios to be concocted to explain the observed correlation: in particular, there are very few theoretical constraints on the strength or geometry of a magnetar's magnetic field. Allowing for plausible variations in the external dipole field, the interior toroidal field, and the geometry at the crust-core boundary $-$ together with uncertainties in the material properties of the crust $-$ could potentially allow one to `explain' each of the sixteen sources we consider here, whilst simultaneously not learning anything about them.

In an effort to harness our observed correlation to probe the underlying physics of magnetar bursting, we will adopt an ansatz that the primary physical parameter that governs bursting activity is simply the magnetar's \emph{true} age. Although this may seem overly restrictive, (i) there is no clear observational evidence that the birth fields of magnetars \emph{do} vary greatly; (ii) simply because a variety of magnetic-field configurations are theoretically permitted, it does not mean they are realized in practice, with e.g., stability considerations being very restrictive; (iii) even if a wide range of birth magnetic fields \emph{are} possible, only a small subset of geometries are likely to lead to magnetar activity \citep{lander25}. As we will see, our ansatz has the virtue of leading to a clear and potentially falsifiable interpretation of the physics driving burst activity $-$ though we cannot, of course, rule out other possible explanations for the burst-activity correlation we find.

\subsection{True age of a magnetar}
\label{true_age}

Figures \ref{fig:energy_vs_age} and \ref{fig:burst_clus_vs_age} show a striking correlation between burst activity and characteristic age. We would like to be able to take the latter quantity as a proxy for \emph{true} age, which would then allow us to draw valuable insights into the evolution of magnetars from these figures. Let us, then, consider how well characteristic and true age are likely to be correlated.

The characteristic age gives a timescale for the present-day spindown of a neutron star, but magnetars can experience substantial spindown variation on short timescales due to their coronal twists \citep{TD01, Thompson2002, Beloborodov2007}. A recent example of this is SGR 1818.0$-$1607, discovered during a period of outburst and described as `very young' with its initially reported $\tau_c=240\,\mathrm{yr}$; but two years later its spindown had stabilized, resulting in a revised (and more typical) $\tau_c=840\,\mathrm{yr}$ $-$ nearly four times the earlier value. The danger of such misleading $\tau_c$ can thus be somewhat ameliorated by avoiding epochs around significant outbursts/bursting activity, and we have done so in the values we report here.

A more striking example of this is SGR 1806$-$20 that showed long-term variability in its spindown rate between 2000 and 2011, with spin frequency derivative ($\dot \nu$) increasing an order of magnitude following its giant flare on 2004 December 27 \citep{Woods2007, Younes2015}. The characteristic age of the source was reported as 0.24 kyr using $P$ and $\dot P$ values measured in this active phase \citep{Olausen2014}, whereas \cite{Younes2017} calculated a characteristic age of 1.6 kyr, based on the observations with NuSTAR after 11 yr following the giant flare, and found that $\dot \nu$ had stabilized to its historical minimum value measured in 1996 by \cite{Woods2000}. However, during the long active phase of the source, its spin period decreased by 3\,\%, far significant than what is expected from the nominal value. Accumulation of this magnitude over several kiloyears could lead to large discrepancies between the characteristic and true age of the magnetar. 

It is occasionally possible to estimate magnetar ages from proper-motion measurements and associations with star clusters; \cite{tendulkar} used this method to estimate the true ages of SGRs 1806$-$20 and 1900+14 as being, respectively, $0.65\pm 0.30\,\mathrm{kyr}$ and $6.0\pm 1.8\,\mathrm{kyr}$ $-$ results are reasonably consistent (within a factor of 5) with the $\tau_c$ we report here. Note that this age-estimation method has its own inherent uncertainties, too. Another possible way to estimate the true age of a magnetar is the calculation of the associated supernova remnant (SNR) age, if any. As an example, PSR 1846.4$-$0258 is located in the young SNR Kesteven 75, whose age is consistent with the characteristic age of the pulsar \citep{Gotthelf2000, Livingstone2011, Leahy2008_psr1846}. On the other hand, 1E 2259+586 with a spindown age of $\sim$\,230 kyr may have a true age of $\sim$\,10 kyr, given its location near the center of the X-ray bright SNR CTB 109 \citep{Sasaki2004, Sasaki2013, Sanchez2018}. This implies that, even at this young age, the evolution can deviate significantly from that of an assumed magnetic dipole (with braking index n=3). The greatest deviations of $\tau_c$ from true age can be expected from sources with $\tau\gtrsim 100\,\mathrm{kyr}$, where evolution of the star's crustal field and rotation mean the long-term average spindown is likely to have been very different from that seen in present-day observations. For example, SGR 0418+5729 has $\tau_c\sim 36000\,\mathrm{kyr}$, but a true age estimated as $\sim 550\,\mathrm{kyr}$ from a magnetothermal evolution \citep{Rea2013}, and just $18\,\mathrm{kyr}$ from a study modeling its magnetospheric evolution \citep{mondal21}. 

Another argument in favor of using Figures \ref{fig:energy_vs_age} and \ref{fig:burst_clus_vs_age} to infer a correlation between burst energy and true age is that there is no obvious physical reason for burst energy to be correlated with $\tau_c$ per se; bursts are believed to be powered by crustal stresses rather than the star's spindown. By contrast, there \emph{is} a good reason to expect burst energy to reduce with true age, as magnetic-field decay results in a significant reduction in magnetic energy over a timescale of $\sim 10\,\mathrm{kyr}$ (see, e.g., \cite{pons_gepp,dehman23}), consistent with \autoref{fig:energy_vs_age}. We explore this in the following subsections, where we argue that the bursting behavior of a magnetar gives us insights into the size of highly-stressed regions in the star's crust and how these change over time.

\subsection{Aging magnetars: the effect of Ohmic decay}

The interplay between Hall drift and Ohmic decay causes the evolution and decay of the crustal magnetic field. To understand this process, we begin by defining the characteristic timescales for these effects:
\begin{equation}
    \tau_{\rm{Ohm}}=\frac{4\pi\sigma L^2}{c^2}\ \ ,\ \ 
    \tau_{\rm{Hall}}=\frac{4\pi\rho_e L^2}{cB},
\end{equation}
where $\sigma$ is the electrical conductivity, $\rho_e$ the charge density, and $L$ a characteristic lengthscale for the magnetic field. We may then approximate the resulting field decay  with the formula \citep{Aquilera2008}
\begin{equation}\label{B_decay}
    B(t)=B_0\frac{\exp(-t/\tau_{\rm{Ohm}})}{1+\frac{\tau_{\rm{Ohm}}}{\tau_{\rm{Hall}}}(1-\exp(-t/\tau_{\rm{Ohm}}))},
\end{equation}
where $B_0$ is the initial field (note that during the phase that interests us, $B(t)$ is insensitive to any choice of $B_0$ greater than $\sim 10^{14}\,\mathrm{G}$, as shown later). From these timescales, we can also define the \emph{Hall parameter}:
\begin{equation}
    R_H\equiv\frac{\tau_{\rm{Ohm}}}{\tau_{\rm{Hall}}}=\frac{\sigma B}{c\rho_e},
\end{equation}
$R_H$ defines the dominant field-evolution mechanism:  $R_H>1$ for Hall drift, $R_H<1$ for Ohmic decay.

Let us first consider the population of magnetars with $\tau_c\gtrsim 5\,\mathrm{kyr}$. Their observed reduction in burst energy cannot simply be that Ohmic decay causes a slowing in stress build-up from Hall drift. Were that the case, then $-$ in the absence of other changes to the crust $-$ we would expect to observe bursts of the same size but reduced frequency as the star ages. Although some evolution of the material properties of the crust is expected $-$ for example, the yield stress does have some dependence on temperature and strain rate \citep{chug_horo}, direction \citep{baiko_kozh}, and crystalline structure \citep{baiko_chug} $-$ these factors alone are unlikely to account for the observed correlation between burst energies and stellar age.

Instead, let us look at the depth of the crustal region where Hall drift is faster than Ohmic decay (i.e., where $R_H>1$), considering a density $\rho$ range from $10^{10}\,\mathrm{g\, cm}^{-3}$ at the top of the crust (a typical choice, avoiding the lowest-density half of the outer crust, which may not even be solid) to $1.4\times 10^{14}\,\mathrm{g\, cm}^{-3}$ at the base, where the crust meets the core. Only when $R_H>1$ do we expect crustal stresses to develop. For young magnetars, with typical field strengths $B\gtrsim 10^{14}\,\mathrm{G}$, the whole crust is in a Hall-dominated state. For a spatially constant crustal $B$, the first region that switches from Hall- to Ohmic-dominated is the base. As a specific example, using the same profiles for $\rho,\rho_e,\sigma$ as in \cite{GL21}, this occurs when $B=1.2\times 10^{13}\,\mathrm{G}$. At progressively lower field strengths, the $R_H<1$ region spreads, leaving an increasingly shallow Hall-dominated layer. Finally, the `top' of the crust reaches $R_H=1$ when $B$ drops to $2.1\times 10^{12}\,\mathrm{G}$. As a result, for a reducing field strength, the maximum elastic energy that can be generated within the crust due to Hall drift also reduces, and when $B\lesssim 2.1\times 10^{12}\,\mathrm{G}$, no `new' stress can be generated anywhere. (Old stresses could, of course, remain from an earlier epoch of Hall drift, unless these have been reduced by previous seismic activity or other effects.)

\begin{figure}[htbp!]
    \begin{center}
    \begin{minipage}[c]{\linewidth}
    \includegraphics[width=\linewidth]{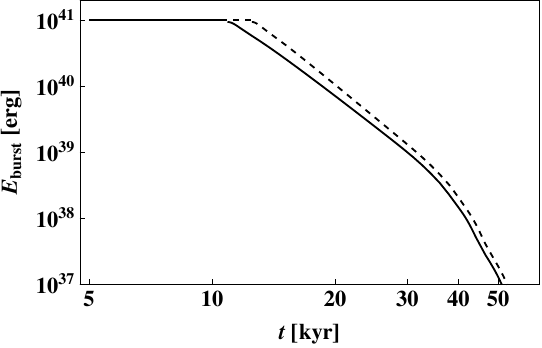}    
    \end{minipage}
    \caption{The reduction in elastic energy storage, due to an encroaching region where Ohmic decay dominates (see text for details), causes the predicted burst energy $E_{\mathrm{burst}}$ to decrease as a function of true age $t$ of a magnetar. Solid and dashed lines show the evolution of models with initial magnetic field strengths of $10^{14}\,\mathrm{G}$ and $5\times 10^{15}\,\mathrm{G}$, respectively.} 
    \label{fig:Eburst}
    \end{center}
\end{figure}

As an illustrative toy model, let us assume that when the entire crust is in a Hall-dominated phase, the largest possible individual short burst is $10^{41}\,\mathrm{erg}$, corresponding to a fixed fraction $1.2\times 10^{-6}$ of the star's maximum possible elastic energy $E_{\rm{max}}$, i.e., the energy that would be stored if the entire crust were at its yield stress. In the gradually-reducing Hall-dominated volume of the crust that results from field decay, the total elastic energy reduces to some fraction of $E_{\rm{max}}$, and the predicted maximum burst energy decreases in the same way. If we now combine this with the prescription for field decay in equation \eqref{B_decay}, we can calculate a predicted burst energy as a function of stellar age, which we plot in \autoref{fig:Eburst}.


\subsection{Very young magnetars}

The brightest short bursts come from sources with $\tau_c$ in the range $0.9-3.6\,\mathrm{kyr}$. Given the uncertainty in how this relates to true age (see section \ref{true_age}), these magnetars might very well all be the same age. Before this broad peak, there are three `young' magnetars, each with a very similar $\tau_c$ value (all  $<0.9\,\mathrm{kyr}$), and each having emitted a large number of relatively weak bursts. Although the weakness of their bursts alone is also consistent with these magnetars being far older than their $\tau_c$ values, the profusion of the bursts indicates that they are indeed young.

Large numbers of weak bursts are consistent with the way stresses are likely to develop in a young magnetar. A magnetar's crust freezes only gradually, from the inside out, and its yield stress is considerably lower when it has just solidified \citep{chug_horo}. Below a true age of $\sim 1\,\mathrm{kyr}$, Hall drift will not have had time to bring a large portion of the crust to its \emph{zero-temperature} yield stress, but frequent failures of its weak, hot, outer crust are plausible. Whilst we should be circumspect about Figure \ref{fig:burst_clus_vs_age}'s apparent positive correlation between burst energy and age for very young ($\tau_c\lesssim 1\,\mathrm{kyr}$) magnetars, given that it is based on a limited number of sources with poorly constrained true age, there is a physical reason to expect such a correlation: progressively deeper failures, releasing more energy, become possible only as the star ages, and its crust cools further and becomes more stressed.

\subsection{Size of burst clusters}

In addition to the correlation between burst energy and $\tau_c$, Figure \ref{fig:energy_vs_age} also shows a notable reduction in the total number of bursts a magnetar is likely to emit as it ages. One might worry that this result could be heavily affected by how long we have been observing a given magnetar, but in fact, magnetars tend to have rare very active episodes (often when they are discovered) $-$ where many bursts are emitted in a cluster $-$ and long periods of quiescence. For example, although the first bursts from SGR 1806$-$20 were seen in 1979 \citep{Cheng1996} and it has been monitored since then, no intense bursting episode has been observed from this source in the last two decades \citep{Gotz2006}. As before, considering the attributes of individual burst clusters allows for a fairer comparison between sources, and the number of bursts in a cluster is seen to decrease with $\tau_c$. Again, taking this as a proxy for true age, it appears that the physical size of active regions on a magnetar decreases notably as it ages.

\begin{figure}[!htbp]
    \begin{center}
    \begin{minipage}[c]{\linewidth}
    \includegraphics[width=\linewidth, trim = 70 260 230 240, clip]{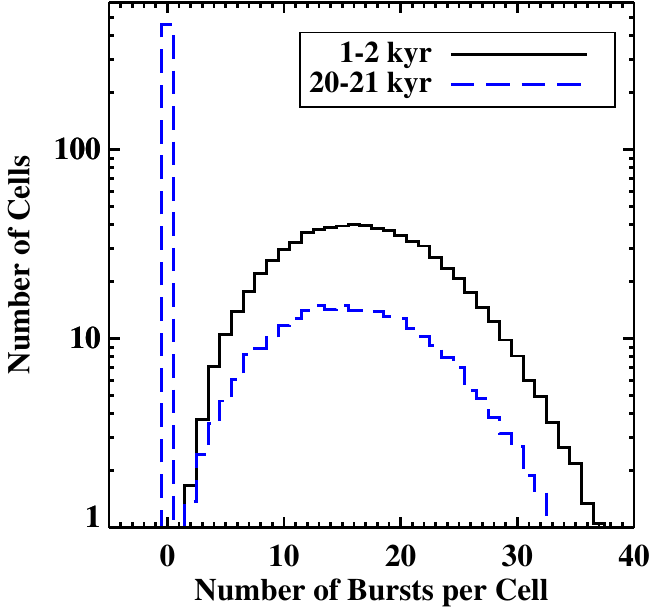}
    \end{minipage}
    \caption{Number of cells vs. number of bursts per cell over 1 kyr evolution.
    The black solid line displays the burst number distribution per cell, obtained as an average of 100 simulations, for a young magnetar whose age is between 1-2 kyr.
    The blue dashed line represents the same relation for an older (20-21 kyr) magnetar.
    The accumulation of cells giving 0 bursts corresponds to cells that have previously suffered deep failures and are therefore currently experiencing no remaining stress.}
    \label{fig:burst_num_hist}
    \end{center}
\end{figure}

Building on our earlier paper \citep{Keskin2024}, where we modeled burst clusters as being due to interactions between local highly-stressed cells of the crust and their neighboring cells, we are also able to interpret the decrease in burst-cluster size with age. In the crust of a young magnetar (the focus of our previous paper), large regions are at a stress close to the elastic yield value, beyond which failure $-$ together with the release of some elastic energy $-$ must occur. When any given cell fails, therefore, it is likely to trigger the successive failure of several neighbors, thus producing a cluster of bursts. An older magnetar, however, may have suffered a number of large events $-$ such as giant flares $-$ over its seismic history, leaving low stress levels in a large volume of the crust. Such extensive failures were excluded from the modeling in \cite{Keskin2024}, which was concerned with short-timescale phenomena, but can trivially be re-enabled in the code, using the original deep-failure prescription from \cite{Lander23}. To understand how burst clustering may change over time, we compare the burst number distribution per cell over 1 kyr of evolution for two scenarios: one young magnetar $-$ assumed to become burst-active at an age of $1\,\mathrm{kyr}$ and not to suffer any deep failures over the following kyr $-$ and one old magnetar that has already evolved over a $20$-kyr period that included deep failures. The two distributions are compared in \autoref{fig:burst_num_hist}. It is clearly seen that cells in the older magnetar are less active, with many being completely inactive (see the accumulation of cells giving 0 bursts in the figure) due to having no remaining stress. As a result, the failure of a cell in an aging magnetar will typically trigger failure in far fewer neighbors, thus reducing the size of burst clusters.


\section{Discussion}\label{sec:discussion}

The suggestion of a correlation between burst activity and magnetar age is not new: \cite{Perna-Pons-2011} found a way to account for this in their pioneering study on the link between crustal failure and magnetic field evolution. What we offer here is a quantitative study of the correlation based on decades of burst data, carefully treated in a way that permits direct comparison. Comparing very young, typical, and aging magnetars, we show how their burst activity helps build a more quantitative picture of how, where, and when crustal stresses develop and are relieved. This complements our earlier work on the burst activity of typical magnetars \citep{Keskin2024}, and on the full range of crust-driven energy output from magnetars \citep{Lander23}.

With clear expectations both from observations and theory, our work provides a method to estimate the age of a magnetar based on its burst activity alone and to construct a complete history of a magnetar's crust. The crust solidifies gradually, from the inside outwards, over a period of centuries. At the point of freezing, it must be in an unstressed state, since a fluid cannot resist shear stresses. This early phase of crustal evolution, taking perhaps $\sim 1\,\mathrm{kyr}$, is complex: as well as undergoing magnetothermal evolution, coupled with plastic flow, the crust itself is changing in both thickness and its mechanical properties. No simulation to date has addressed this in a self-consistent manner. However, given the newborn crust's limited ability to develop and release significant stress, we predict that no `conventional' magnetar activity $-$ i.e., powered by crustal stresses $-$ will be detectable from a star with age $\lesssim 0.5\,\mathrm{kyr}$.

As a magnetar approaches an age of $\sim 1\,\mathrm{kyr}$, crustal failures have the potential to become progressively deeper and to release more energy. The number of bursts in a typical cluster is large; any local failure is likely to trigger neighboring failures. Magnetars in the age range $\sim 1-10\,\mathrm{kyr}$ are likely to display fairly similar burst activity, both in energy and number. This is, however, also the phase where very rare giant flares are most likely to occur, releasing huge amounts of energy and destressing large volumes of the crust \citep{Lander23}.

When the low-stress regions have grown sufficiently in volume, the magnetar's ability to produce large burst clusters becomes compromised. Furthermore, new stresses can develop only within a diminishing thickness of the crust, reducing the average burst energy. This phase, for an age $\sim 10-100\,\mathrm{kyr}$, ends with the magnetar becoming silent.

In this study, we chose to analyze the correlation of magnetar burst activity with characteristic age, as calculated from the observed spindown parameters $P$ and $\dot{P}$. We could instead have compared burst activity with inferred dipole field strength, which is also derived from spindown parameters, and is slightly less sensitive to fluctuations in $\dot{P}$. It is, however, a more indirect proxy for the age of the source. A plot showing the correlation of burst activity with inferred dipole field strength is given in Appendix \ref{appC}.

The scenario we propose in this paper is, of course, based on qualitative results about field evolution rather than the detailed 3D magnetothermal evolutions of the problem that are now available, which couple both the magnetic-field and thermal evolution of the crust, with realistic microphysical parameters (e.g., \cite{degrandis21,igoshev21,ascenzi24}). What these studies miss, however, is a realistic treatment of crustal failure, and so can only be taken as truly quantitative when the crust is below its yield stress. Furthermore, the initial field and the crust-core boundary condition used in simulations can lead to serious variations in the evolution and predicted burst activity \citep{dehman23,lander25}. These uncertainties provide a rationale for more qualitative work like ours, focusing on interpreting observations through models of the build-up and release of stress, with the intention that this can guide more realistic treatments of crustal evolution at high stress in the future.


\begin{acknowledgments}
This work was supported by a Royal Society International Exchange grant IES$\backslash$R3$\backslash$223220 between the University of East Anglia and Sabanc\i\ University.
\end{acknowledgments}


\appendix

\section{THE SAMPLE} \label{appA}

We list our sample in \autoref{tab:MagnetarList}. The list includes the name of the sources, their spin periods ($P$) and period derivatives ($\dot P$), inferred dipolar magnetic field strengths ($B$) and characteristic ages ($\tau_c$) by using $P - \dot P$, as well as their distance estimates (D) and the energy ranges of their emitted short bursts, together with their all corresponding references.

\section{Distances} \label{appB}

To calculate the burst energetics of the sources in our sample, we utilized the distances to the sources provided and commonly accepted in the literature. Many of the sources in our sample were assigned to a distance with large uncertainties by associating them with a region, a nearby system, a star cluster, or a supernova remnant (SNR). These uncertainties may stem from measurement/method errors or approximate estimations.

As examples of magnetars that are associated with a region, SGR 0418+5729 and SGR J0501+4516 both lie on the galactic plane and in the direction of galactic anticenter \citep{van2010}. Since many stars in that direction reside in the Perseus spiral arm of the Milky Way, and the distance to the arm was determined to be 1.95$\pm$0.04 kpc by \citep{Xu2006}, the distance to both magnetars was commonly accepted as $\sim$\,2 kpc in the literature. SGR J1745$-$2900 serves as an example of a source whose distance is obtained due to its proximity to a known system: It sits near the Galactic center (Sagittarius A*), which is located at a distance of 8.3$\pm$0.3 kpc from the Sun \citep{Bower2014}. Similarly, SGR J1833$-$0832 is assumed to be at a distance of 5.67 kpc, based on a likely association with the nearby star-forming region containing PSR J1830$-$08, which is roughly 3$'$.7 away from the magnetar \citep{EG10}.

One alternative way of calculating distances is using spectroscopy for the associated host clusters that contain magnetars. \cite{Bibby2008} and \cite{Davies2009} used this technique and obtained the distances to the magnetars, SGRs 1806$-$20 and 1900+14, and their associated clusters as $8.7^{+1.8}_{-1.5}$ kpc and 12.5$\pm$0.3 kpc, respectively. Magnetar$-$SNR associations provide another alternative way for distance measurements. For example, SGR J1550$-$5418 is associated with the radio shell G327.24$-$0.13. By associating these with SNRs G326.96+0.03 and G327.99$-$0.09 that reside within the distance range of 3.7$-$4.3 kpc, \cite{Gelfand2007} proposed a distance of $\sim$\,4 kpc to SGR J1550$-$5418. Using the observations of the X-ray rings centered on the same magnetar following the burst storm observed on 2009 January 22, \cite{Tiengo2010} measured the distance as $\sim$\,4$-$5 kpc, consistent with \cite{Gelfand2007}. 1E 2259+586 is another source that is associated with an SNR, CTB 109, whose location is assumed to be within or close to the Perseus arm spiral shock at a distance of 3.2$\pm$0.2 kpc \citep{Kothes2012}. However, for the same source, using red clumps which are core helium-burning stars and serve as good infrared standard candles, \cite{Durant2006} estimated the distance as 7.5$\pm$1.0 kpc. Finally, the sources 1E 1841$-$045 and PSR 1846.4$-$0258 are associated with SNRs Kesteven 73 and Kesteven 75, whose distances to us were found to be $8.5^{+1.3}_{-1.0}$ kpc \citep{Tian2008} and $6.0^{+1.5}_{-0.9}$ kpc \citep{Leahy2008_psr1846}, respectively.

\section{Magnetic Field} \label{appC}

\begin{figure*}[htbp!]
    \centering
    \includegraphics[width=0.75\textwidth, trim=45 260 60 235, clip]{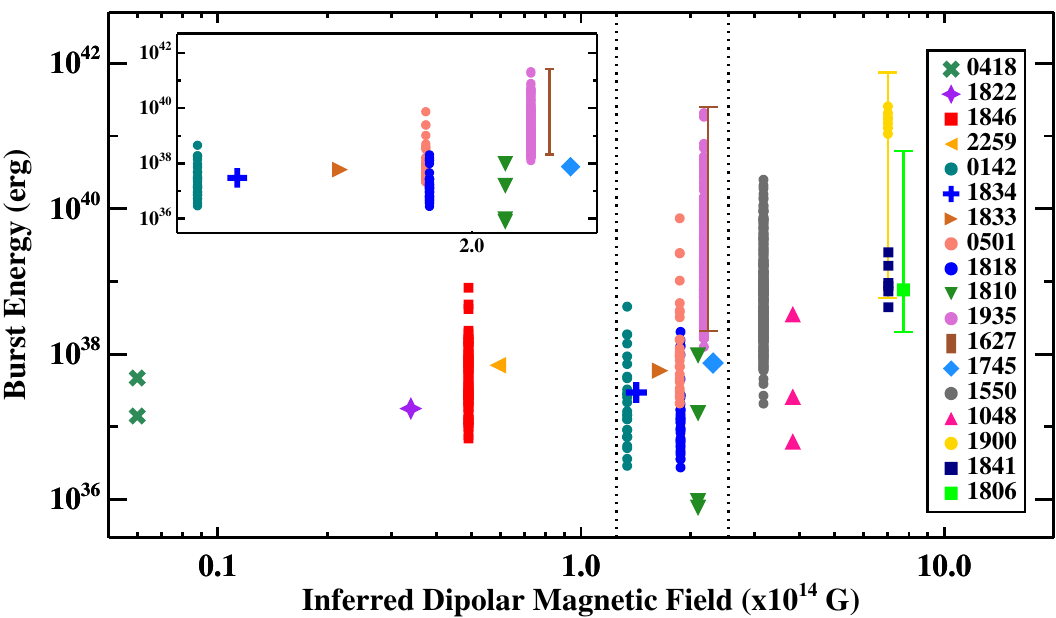}
    \caption{Plot of burst energies vs. inferred dipolar magnetic field strengths of burst active magnetars. Filled symbols represent the energetics of individual bursts from the color-coded magnetars. For SGR 1900+14 (gold), SGR 1806$-$20 (green), and SGR 1627$-$41 (brown), we present burst energy ranges rather than individual burst energies (see text for details). Symbols on the right are ordered in increasing inferred dipolar magnetic field strength. The inset zooms in on the region that remains between the dotted lines.} 
    \label{fig:energy_vs_bfield}
\end{figure*}

We present magnetars' burst energies vs. their inferred dipolar magnetic field strengths in \autoref{fig:energy_vs_bfield}. Note again that for the activities observed with \cgro and \integral from the sources, SGR 1900+14 (gold), SGR 1806$-$20 (green), and SGR 1627$-$41 (brown), we present burst energy ranges. For details of the observations and the calibration of burst energetics, see Section \ref{sec:observation}.


\begin{deluxetable*}{lccccccr}
\tablecaption{List of Burst Active Magnetars and Their  Characteristics}
\tablewidth{0pt}
\tablehead{
\colhead{\bf Source Name} & \colhead{\bf P} & \colhead{ $\dot{\bf P}$}  & \colhead{ $\bf \tau_c^a$} & \colhead{\bf B$^{b}$} & \colhead{\bf D} & \colhead{\bf Burst Energy$^{c}$} & \colhead{\bf Reference$^e$} \\
 & \colhead{\bf (s)} & \colhead{\bf (10$^{-11}$ s s$^{-1}$)}   & \colhead{\bf (kyr)}  & \colhead{\bf (10$^{14}$ G)} & \colhead{\bf (kpc)}  & \colhead{\bf (10$^{38}$ erg)} & } 
\startdata
\hline
AXP 4U 0142+61 & 8.69 & 0.20 & 68 & 1.34 & 3.6 & [0.03, 4.5] & (1), (2), (3)\\
\hline
SGR 0418+5729 & 9.08 & 0.0004  & $\sim$36,000 & 0.06 & 2 & [0.14, 0.47] & (4), (5), (6)\\
\hline
SGR 0501+4516 & 5.76 & 0.594 & 15 & 1.87 & 2 & [0.21, 73.72] & (7), (8), (9)\\
\hline
1E 1048.1$-$5937 & 6.46 & 2.22 & 4.6 & 3.83 & 9 & [0.06, 3.53]$^{\bf d}$ & (1), (2), (10, 11)\\
 \hline
 SGR J1550$-$5418 & 2.07 & 4.77 & 0.69 & 3.18 &  5 & [0.21, 250.56] & (12), (13), (6)\\
 \hline
SGR 1627$-$41 & 2.59 & 1.90 & 2.2 & 2.24 & 11 & [2.07, 2532.66] & (14, 15), (16), (17)\\
\hline
SGR J1745$-$2900 & 3.76 & 1.39 & 4.3 & 2.31 & 8.3 & 0.75 & (18), (19), (20)\\
\hline
SGR 1806$-$20 & 7.75 & 7.51 & 1.6 & 7.72 & 8.7 & [2.02, 619.31] & (21), (22), (23, 24, 6)\\
\hline
XTE J1810$-$197 & 5.54 & 0.78 & 11 & 2.10 & 3.5 & [0.008, 0.95]$^{\bf d}$ & (25), (26), (27)\\
\hline
Swift J1818.0$-$1607 & 1.36 & 2.55 & 0.84 & 1.88 & 4.8 & [0.03, 2.02] & (28), (29), (30)\\
\hline
Swift J1822.3$-$1606 & 8.44 & 0.013 & $\sim$1000 & 0.34 & 1.6 & 0.18 & (31), (32), (6)\\
\hline
SGR J1833$-$0832 & 7.57 & 0.35 & 34 & 1.65 & 5.7 & 0.59 & (33), (20), (34)\\
\hline
Swift J1834.9$-$0846 & 2.48 & 0.796 & 4.9 & 1.42 & 4.2 & 0.30 & (35), (36), (37)\\
\hline
1E 1841$-$045 & 11.78 & 4.08 & 4.6 & 7.02 & 8.5 & [4.41, 25.07] & (1), (38), (6)\\
\hline
PSR J1846.4$-$0258 & 0.33 & 0.71 & 0.74 & 0.49 & 6 & [0.07, 8.16] & (39), (40), (30)\\
\hline
SGR J1900+14 & 5.20 & 9.20 & 0.9 & 7.00 & 12.5 & [5.95, 7432.91] & (41), (42), (43, 44)\\
 \hline
 SGR J1935+2154 &  3.24 & 1.43 & 3.6 & 2.18 & 9 & [1.26, 2065.91] & (45), (46), (47)\\
 \hline
 1E 2259+586 & 6.98 & 0.048 & $\sim$230 & 0.59 & 3.2 &  0.70 & (1), (48), (6)\\
\enddata
\tablecomments{\\
$^{a}$ Characteristic Age, $\tau_c = P / 2\dot{P}$ \\
$^{b}$ Inferred dipolar magnetic field strength, $B = (3 c^3 I P \dot P / 8 \pi^2 R^6)^{1/2}=3.2 \times 10^{19} (P \dot P)^{1/2}$ assuming $I$ = 45 g\,cm$^2$ and $R$ = 10 km\\
$^{c}$ In 8$-$200 keV.\\
$^{d}$ In 2$-$200 keV.\\
$^{e}$ Numbers given in parentheses correspond to the references of $P \& \dot P$, distance ($D$), and burst energetics for each source, respectively.\\
References: (1) \cite{Dib2014}; (2) \cite{Durant2006}; (3) \cite{EG17}; (4) \cite{Rea2013}; (5) \cite{van2010}; (6) \cite{Collazzi2015}; (7) \cite{Camero2014}; (8) \cite{Xu2006}; (9) \cite{Lin2011}; (10) \cite{Gavriil2002}; (11) \cite{Gavriil2006}; (12) \cite{Dib2012}; (13) \cite{Tiengo2010}; (14) \cite{Espocito2009}; (15) \cite{Esposito_2009}; (16) \cite{Corbel1999}; (17) \cite{Wood1999}; (18) \cite{Kaspi2014}; (19) \cite{Bower2014}; (20) \cite{Kennea2013}; (21) \cite{Younes2017}; (22) \cite{Bibby2008}; (23) \cite{EG20}; (24) \cite{Gotz2006}; (25) \cite{Camilo2007}; (26) \cite{Minter2008}; (27) \cite{Wood2005}; (28) \cite{Rajwade2022}; (29) \cite{Lower2020}; (30) \cite{Uzuner2023}; (31) \cite{Castillo2016}; (32) \cite{Scholz2012}; (33) \cite{Esposito2011}; (34) \cite{EG10}; (35) \cite{Kargaltsev2012}; (36) \cite{Leahy2008_sgr1834}; (37) \cite{Delia2011}; (38) \cite{Tian2008}; (39) \cite{Livingstone2011}; (40) \cite{Leahy2008_psr1846}; (41) \cite{Mereghetti2006}; (42) \cite{Davies2009}; (43) \cite{EG99}; (44) \cite{Israel08}; (45) \cite{Israel2016}; (46) \cite{Lin2016}; (47) \cite{Lin2020}; (48) \cite{Kothes2012}
}
\label{tab:MagnetarList}
\end{deluxetable*}



\clearpage

\bibliographystyle{aasjournal}
\bibliography{refs}

\end{document}